**NMR Spectroscopy Can Help Accelerate Antiviral Drug Discovery Programs**

Steven R. LaPlante,*[1,2,3] Pascale Coric[3], Serge Bouaziz[3], Tanos C. C. França[1]

[1] Pasteur Network, INRS-Centre Armand-Frappier Santé Biotechnologie, 531 Boulevard des Prairies, Laval, Québec, H7V 1B7, CANADA
[2] NMX Research and Solutions, Inc., 500 Boulevard Cartier Ouest, Laval, Québec, H7V 5B7, CANADA
[3] Université Paris Cité, CNRS, CiTCoM, F-75006 Paris, France

* Corresponding author: steven.laplante@inrs.ca

**Abstract**

Small molecule drugs have an important role to play in combating viral infections, and biophysics support has been central for contributing to the discovery and design of direct acting antivirals. Perhaps one of the most successful biophysical tools for this purpose is NMR spectroscopy when utilized strategically and pragmatically within team workflows and timelines. This report describes some clear examples of how NMR applications contributed to the design of antivirals when combined with medicinal chemistry, biochemistry, X-ray crystallography and computational chemistry. Overall, these multidisciplinary approaches allowed teams to reveal and expose compound physical properties from which design ideas were spawned and tested to achieve the desired successes. Examples are discussed for the discovery of antivirals that target HCV, HIV and SARS-CoV-2.

**Keywords**: Drug discovery, NMR, antivirals, drug design, lead discovery, infectious diseases, microbes, virus, fragment screening, inhibitor

# 1. NMR roles in supporting multidisciplinary efforts to reveal compound properties that helped to generate drug design ideas.

1.1 Drug design, revealing ligand-protein interactions, and optimizing properties.

The central goal of all antiviral drug discovery programs is to discover and optimize compounds that kill viruses but are also safe for human consumption. Considerable efforts are therefore made, and procedures developed, to ensure that compounds are designed to have the appropriate properties to achieve these desired effects.

This review describes hands-on efforts and procedures developed to discover inhibitors that targeted the critical proteins of hepatitis C (HCV), SARS-CoV-2 and the human immunodeficiency (HIV) viruses. In particular, we focus on the applications and roles of NMR spectroscopy in achieving those goals. Salient examples are described where NMR applications ranged from library screening to the design of antiviral drugs when strategically combined with multidisciplinary techniques.

Central to these successes was defining the appropriate and feasible NMR experiments that could properly reveal relevant compound and protein properties, from which rational strategies were made to stepwise advance antiviral programs. For the sake of brevity, some NMR experiments are listed below from which key ligand and protein properties (underlined for easy visual recognition) were reliably extracted.

First, an important question addressed via NMR was to <u>determine whether or not ligands bound directly to target proteins.</u> Although this question appeared to be simple, it was noteworthy that orthogonal biophysics techniques often disagreed and thus caused confusion. NMR had significant advantages for addressing this simple question given that the protein and ligands are free in solution and untethered to any type of solid support. Furthermore, automation tools enabled NMR experiments to be practical for screening libraries of thousands of compounds or simply for validating hits derived from other library screening strategies. Scheme 1 showed a cartoon of how NMR was used to distinguish between a ligand that did not bind to a target protein (left, non-binder) versus a ligand that selectively and stoichiometrically bound to the target protein (center, Binder). Orthogonal ligand-detection NMR methods were employed to make these distinctions – $^1$H and $^{19}$F differential line broadening (DLB), Carr-Purcell-Miboom-Gill relaxation (T2-CPMG), WATER-LOGSY and/or saturated transfer difference (STD). Note that these experiments work well for detecting ligand binding (e.g. single-digit μM Kd and weaker) which were in fast- and intermediate-exchange (on the NMR timescale) between the free and bound states. The detection of more potent ligand binding (e.g. $K_D$ in the nM range) that were in slow-exchange (on the NMR timescale) between the free and bound states, were observed by ligand-detection 19F NMR or by protein-detection HSQC experiment.

Protein-detection methods were employed to identify ligand binding (for fast, intermediate and slow exchange systems) via chemical shift perturbations in heteronuclear single-quantum correlation $^1$H-$^{15}$N/$^{13}$C HSQC spectra. These experiments along with 1D 1H/19F NMR competition experiments were also used to specifically identify ligands binding pockets on target proteins.

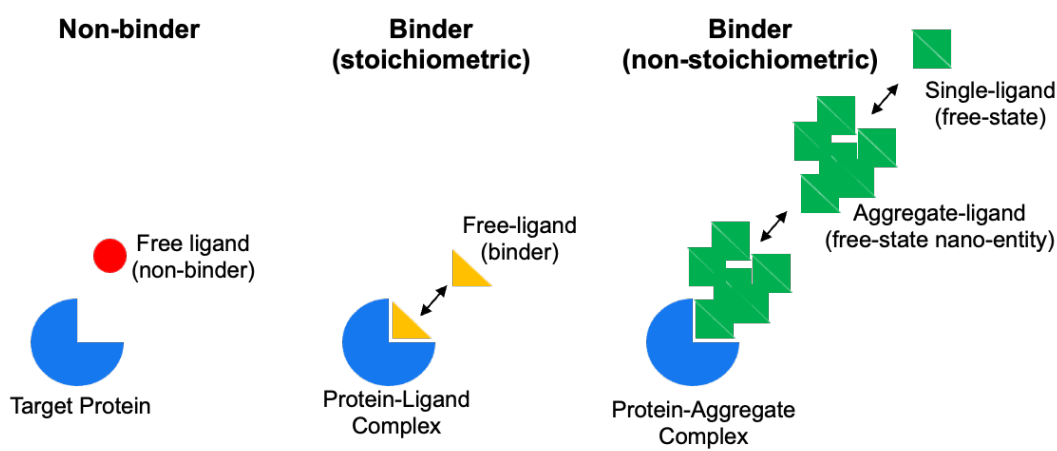

**Scheme 1:** Shown is a cartoon of three distinct ligand-protein properties that can be detected using NMR spectroscopy.

Other critical properties were easily exposable by NMR strategies. For example, knowledge of the free-state solution properties of compounds helped to determine solubility, structural integrity and bioactive conformational flexibility/rigidity. One-dimensional $^1$H (1D $^1$H) NMR, spin-lattice relaxation ($^{13}$C-$T_1$), and $^1$H-$^1$H J-couplings were valuable for this. Compound chirality that resulted from atropisomerism and conformer exchange were identified by 1D $^1$H, ROESY and variable temperature experiments (VT). NMR was also essential in detecting properties that frequently resulted in false-positives and wasted chemical optimization efforts such as compound self-aggregation into nano-entities [20-23,46]. Scheme 1 displays this on the right side. To detect this property, methods were developed that used one-dimensional $^1$H (1D $^1$H) and CPMG NMR methods.

The bioactive target-bound conformations of ligands were deciphered by NOESY experiments. Whereas differences between ligand free- vs bound-states were exposed by comparison of ROESY, transferred NOESY, $^1H$-$^1H$ J-coupling and $^{13}C$-$T_1$ and transferred $^{13}C$-$T_1$ experiments. Design ideas that minimized the ligand entropic costs upon binding first required the identification of sites that conformationally rigidified once attached to the target protein and used $^{13}C$-$T_1$, transferred $^{13}C$-$T_1$, ROESY and $^1H$-$^1H$ J-couplings data. Ligand epitope mapping efforts helped to identify solvent-exposed versus pocket-bound subunits of ligands – using mainly DLB experiments.

This, along with the other strategies, helped to elucidate structure-activity-relationships (SAR) when combined as a multidisciplinary approach. The stoichiometry and relative specificity of proteins and ligands were evaluated by titration experiments involving 1D $^{19}F$ and $^1H$ titration experiments along with 2D HSQC data. The affinities ($K_D$) and affinity ranking of ligands binding to target proteins were determined by titration experiments via HSQC along with 1D $^1H$ and 1D $^{19}F$ NMR, and by scoring changes in STD, DLB, CPMG experiments.

Binding sites and mechanisms of inhibition could be deciphered via HSQC, competition experiments, $^{19}F$ NMR and NOESY information. Target dynamics and tumbling attributes, along with protein folding versus unfolding were exposed by HSQC, $T_1$, NOE and 1D $^1H$ experiments. Finally, exposing attributes related to contamination from false-positives, ligand toxicity, immune responses and promiscuity are also discussed herein.

1.2. Initiating drug discovery programs (hit finding) and the role of NMR

All drug discovery programs must start by identifying an initial hit compound that can then be optimized toward more advanced compounds which have the desired properties (leads). However, there are a limited number of strategies that can be employed for discovering those initial hits. In this review, the role of NMR is discussed for supporting each multidisciplinary strategy in a variety of hit-to-lead projects. The discussions pay particular attention to the fact that NMR biophysics techniques must be employed strategically and pragmatically within team workflows and timelines.

Different strategies have been developed to discover hits. A commonly applied approach for hit discovery is "patent busting", and it is often referred to as a "me too" strategy. An example of this is discussed in Section 2.1. Drug repurposing is regaining more popularity and an example is given in Section 5. High-throughput screens have been highly successful, and applications are mentioned in Sections 2.3, 4.1, 4.2 and 4.4. The screening of DNA encoded libraries (DEL) is becoming a very attractive method; however, it will not be covered in this review. Fragment-based lead/drug discovery (FBLD/FBDD) is briefly discussed in Section 2.2, and a new method is shown in Section 5 that combines FBLD and drug repurposing. Phenotype screening also has been successful for discovering leads. It will not be covered in this review; however, it is mentioned briefly in Section 5. The exploitation of natural products to produce peptidomimetics is shown in Sections 2.1 and 4.3.

## 2. Design of antiviral drugs that target HCV

The orchestration of an effective antiviral campaign can be very complicated, especially when one must also consider the necessity of combating the emergence of viral resistance. Thus, the strategies discussed in this review give examples of inhibiting several viral protein targets as a means to prepare the potential use in combination therapies that have proven to be effective.

2.1. Targeting HCV protease with peptidomimetics starting from a natural peptide substrate.

The first viral protein target to be discussed here is the NS3 serine protease of HCV [1]. Efforts began with "knowledge building" to better understand the characteristics of this target as a potential source of ideas. When the hepatitis C virus enters a human liver cell, it employs its own genetic material and engages the machinery of the cell to prepare for viral replication. A single-chained viral polyprotein is produced that then associates with the cell's endoplasmic reticulum, which then undergoes requisite polyprotein cleavage into individual viral proteins [1] (Fig 1A). Real-time NMR was used to monitor the cleavage of a consensus peptide that corresponded to those cleave sites, and discovered that post-cleavage, the N-terminal product peptide DDIVPC remained bound to the protease and inhibited the protease with a weak inhibition constant of $IC_{50}$ of 71 µM (Fig 1B) [2, 3]. This surprise finding led to the initiation of the design of peptidomimetics based on DDIVPC that contained a C-terminal carboxylic acid.

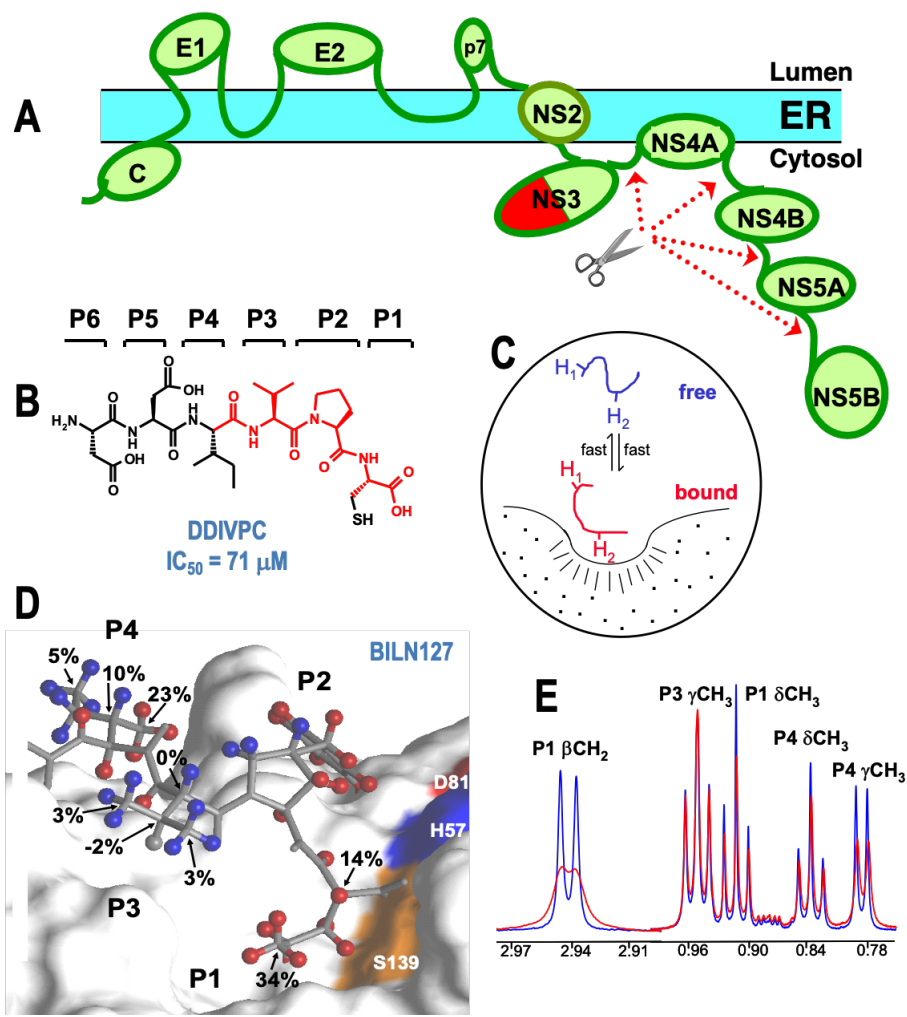

**Figure 1.** Targeting HCV protease. (A) Cartoon of the single-chain HCV polyprotein associated with the endoplasmic reticulum, along with the NS3 cleavage sites. (B) Consensus amino acid cleavage sequence of HCV NS3 serine protease. (C) Requirement of many NMR methods to prove that the ligand and protein bind in a fast-exchange on the NMR time-scale. (D) NMR transferred NOESY structure of BILN127 docked to apo HCV protease. Hydrogens are colored red for which a strong DLB was observed, and colored blue for hydrogens that experience a minor or no DLB. Percentage values near the respective hydrogens were determined by the changes in $^{13}C$ $T_1$ data before and after adding protease [7], (E) DLB results from an overlay of the NMR spectrum of free BILN127 (blue) and BILN127 after adding protease (red) [5].

Contributions from further "knowledge building" again became crucial, especially since there was a lack of available X-ray crystal structures, and multiple attributes and properties were needed to support medicinal chemistry efforts. Given the peptide's fast-exchange binding process on the NMR timescale (Fig. 1C), transferred NOESY experiments allowed for the determination of the bioactive conformation of the peptide when bound to the protease (Fig 1D) [4, 5]. This rigid-body NMR structure was then docked to the active site of the X-ray-derived structure of apo HCV protease. This NMR-derived complex revealed that the end-terminal product peptide DDIVPC bound in the extended conformation as shown in Fig 1D.

A multitude of NMR experiments exposed other properties that enhanced our understanding, and which led to design ideas. For example, 1D NMR experiments of free *versus* bound spectra allowed for epitope mapping to identify which parts of DDIVPC (and other analogues including BILN127) where solvent exposed *versus* those that came into direct binding contact into pockets or on surfaces of the protease. Fig 1E showed that NMR peaks of specific hydrogens of the free ligand (P1 $\gamma CH_2$ displayed in blue) changed dramatically upon addition of small amounts of HCV protease (red), therefore reporting that this group bound directly into a pocket. On the other hand, the NMR peaks of P3 $\gamma CH3$ did not change upon addition of HCV protease (compare the blue and red peaks in the overlay), thus this group did not engage with a pocket and was solvent exposed in the free and bound states. Fig. 1D summarized the DLB data where hydrogens that experienced major changes are colored red, and those that did not were colored blue. Overall, the DLB data were very well consistent with the docked bioactive conformation and complex shown in Fig. 1D. For example, the structure showed that the P1 segment lied within the shallow S1 pocket of the protease and the P3 segment was solvent exposed [4, 5]. Another example was the DLB observation that P5 and P6 were solvent exposed and unstructured in the free and bound states (data not shown) and thus could be subsequently removed to form truncated peptidomimetic analogues as shown in Figure 2 [6].

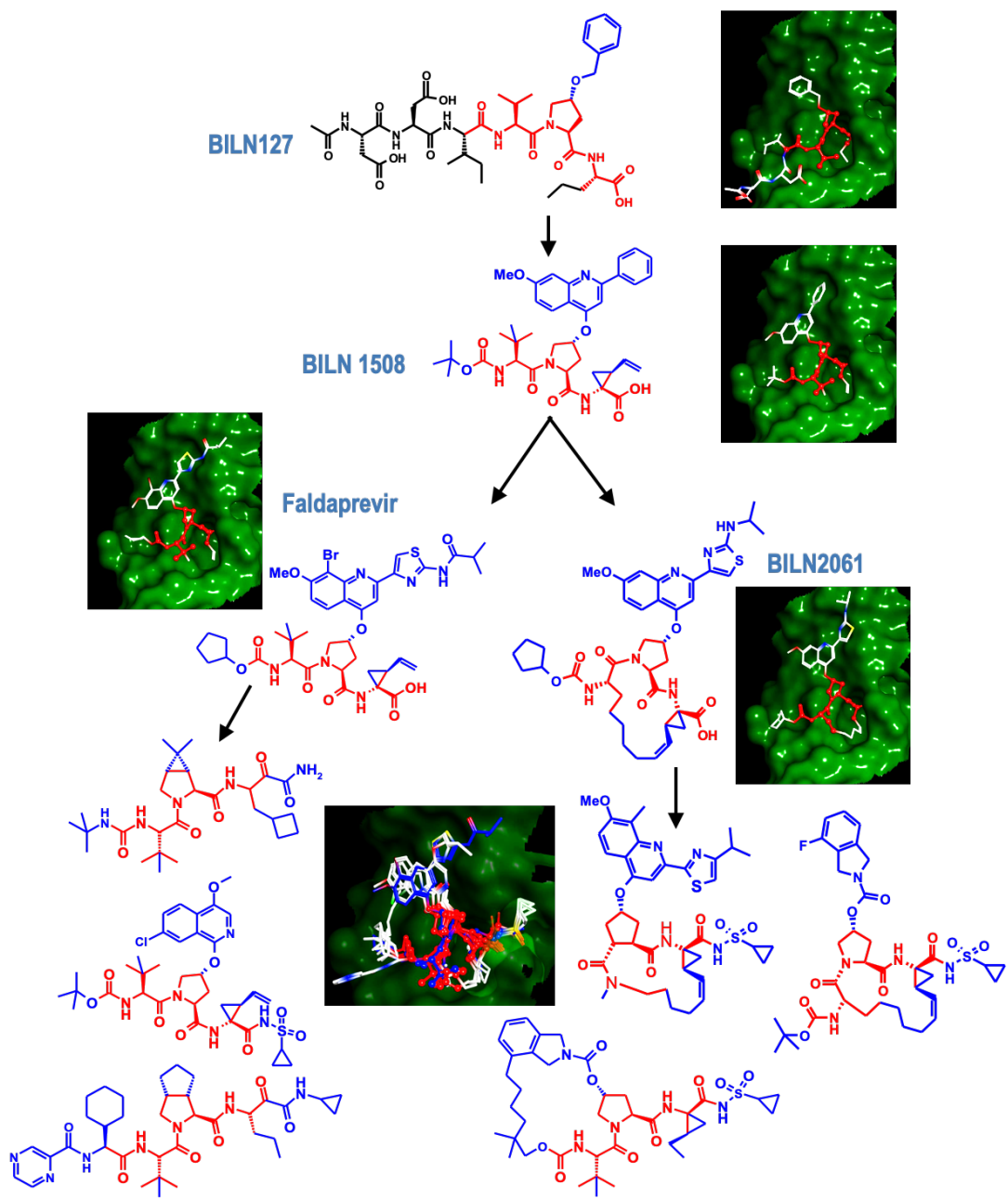

**Figure 2.** Targeting HCV protease. Shown are inhibitors of HCV protease along with X-ray and docked complexes.

Further information was sought regarding differences between the free-and bound states conformation and dynamics of compounds. The aim was to identify any differences and then propose design ideas. Comparisons were made between ROE distance information (from ROESY data) of the free-state with the NOE data (from transferred NOESY) of the bound state. It was notable that both reported similarities consistent with extended backbone conformations in both states [2,4]. However, the comparisons also reported that the sidechains differed significantly. It was then found that these differences were due to flexibility of the sidechains in the free-state and rigidification to the bioactive conformation upon binding [2,4]. To better characterize this, a new, site-specific $^{13}$C $T_1$ experiment was created and called the transferred $^{13}$C $T_1$ which identified segments that rigidified upon binding and those that do not [7]. A summary of the changes in $^{13}$C $T_1$ data upon binding was displayed as percentage values next to each carbon in Fig. 1D. Note that the P1 values were high indicating that the P1 segment seriously rigidified upon binding, whereas the P3 did not as judged by its near zero values. Thus, the entropic cost was high for P1 binding. A breakthrough idea then came to light given this data along with the fact that P1 and P3 were close in proximity in the bound state (Fig 1D). As a means to rigidify the critical P1 segment to resemble the bioactive conformation, a cyclopropyl group at the P1 was made and a macrocycle chemically linked P1 side-chain to the P3 anchor [5].

Thus, it became clear that NMR-based "knowledge building", when used in combination with other disciplines, can prove to be very useful for drug design purposes. Although detailed presentations were given elsewhere, Fig. 2 provided a summary of key compounds that resulted from this chemical evolution. As a brief summary, non-natural amino acids were scanned at P2 and it was found that proline derivatives with the blue-colored groups provided potency advantages (see BILN127, BILN1508 and others in Fig. 2). The black-colored P6, P5 and P4 segments were truncated (see BILN1508 in Fig 2), and modifications were made to the blue-colored P4 segments [2-6, 8].

The structure of BILN2061 illustrated the macrocyclization from the P1 to P3 segments, and this compound was published as the first direct-acting small-molecule antiviral to inhibit hepatitis C in infected patients [4, 5, 9]. This, and other analogues such as BILN1508 and Faldaprevir, caught the attention of many pharmaceutical companies which prompted a flood of "me too" patent-busting strategies. See the examples of alternate drugs at the bottom part of Fig. 2 [4]. It is interesting to note that after the design of tens of thousands of these peptidomimetics by many pharmaceutical companies, the red-colored segments in Fig 2 showed the common and essential features (i.e. the main backbone portions) that stayed constant from the initial hit DDIVPC to all of the effective drugs. Thus, this represents an important lesson for peptidomimetic design in general.

2.2. Targeting HCV helicase with compounds derived from NMR fragment-based lead discovery (FBLD)

Further efforts were also focused on discovering inhibitors of HCV NS3 helicase. To secure lead compounds as starting points, many approaches such as high throughput screening were tried and failed. Many consider HCV helicase as an undruggable protein. However, we found that fragment-based lead/drug discovery (FBLD/FBDD) provided fruitful leads [10]. Although multiple biophysical detection methods were employed for FBLD, NMR was considered as the most practical and successful. For example, NMR could be used to screen unanchored protein targets (unlinked to a plate which is required for SPR and MST, etc.), and NMR was very sensitive for detecting ligands that exhibit very poor affinity binding to the target protein, which is a typical hallmark of hits from a fragment-based screening campaign.

The FBLD approach involved the NMR screening of a library of small fragment-like compounds to detect direct binding to the target NS3 helicase protein (Fig 3B) [10]. Once identified, hits were then confirmed for selectivity and stoichiometric binding attributes. The hit compound in Fig 3A provided an interesting scaffold which had a weak $IC_{50}$ of 500 μM. Once confirmed, a multidisciplinary approach was employed where medicinal chemists synthesized related analogues to help drive efforts along with support from NMR validation studies. Subsequent analogues consisted of adding appendages to the core as a "scaffold growing" strategy to rationally design new compounds that better fit into helicase receptor pockets which resulted in more potent compounds (Fig. 3C). X-ray structure of the complex (Fig. 3D) helped to evaluate the potential new interactions at the atomic level.

More recently, our team introduced a new application of "NMR for SAR" that determined high-throughput affinity ranking of related analogues [11]. This technique is superb for enabling medicinal chemists to establish essential SAR. Interestingly, the example published detected an initial screening hit of 12 mM, and then applied NMR affinity ranking to design sub-micromolar compounds. This work clearly showed how NMR can be central for driving drug discovery projects, and can be valuable when integrated within multidisciplinary efforts.

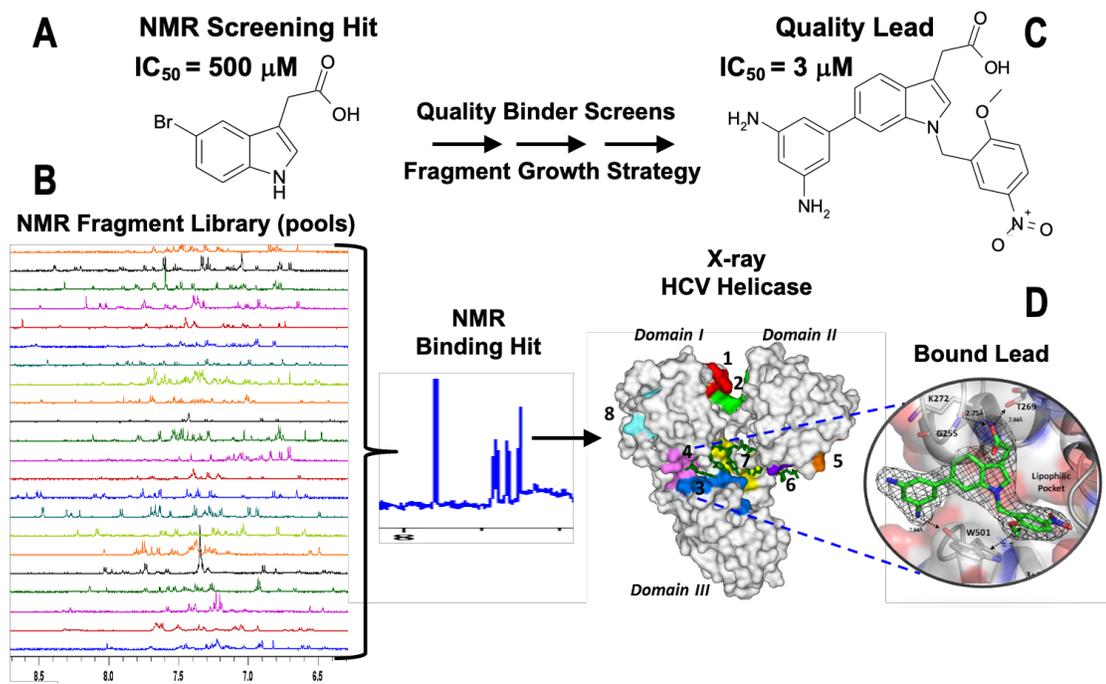

**Figure 3.** Targeting HCV helicase. (A) NMR screening hit. (B) Example of NMR data from a fragment screen. (C) Lead compound derived from the NMR screening hit. (D) X-ray structure of the complex involving the lead in (C) and HCV helicase.

2.3. Targeting HCV polymerase with compounds derived from a high-throughput screening campaign.

Another important method for discovering lead compounds is via high-throughput screening (HTS) campaigns. Historically, this has been one of the most utilized techniques by the pharmaceutical industry, however, alternatives are more recently becoming utilized.

To identify inhibitors of HCV polymerase, a robust biochemical assay was setup that tested for HCV polymerase activity [12, 13]. This assay was then utilized to screen a large collection of over a million compounds to reveal inhibitors. However, an important issue encountered was that too many hits were found. Typical of HTS campaigns, the screen was contaminated with many false-positive hits due to the phenomenon of compound self-association into aggregates (to be discussed in Section 3). Given this, it was impossible to apply follow-up synthesis of analogues on all hits. To reduce the number of hits, multiple counter screens were introduced. Counter screens involving assays with polymerase that were unrelated to HCV served to detect and filter out hits that were non-specific for HCV polymerase activity [12,13]. These counter screens were successful in reducing the number of hits, and the resultant hundreds of hits were then subjected to NMR assays aimed at identifying compounds that bound directly to HCV polymerase [12] and deprioritize those hits that also self-associated and formed nano-entities in the free-state.

The NMR assay identified the hit compound shown in Fig. 4A, which eventually served as the seed for discovering the clinical compound Deleobuvir (Fig. 4E) [14]. The NMR assay compared the $^1$H NMR spectra of the free compound versus that of the compound in the presence of small amounts of HCV polymerase (Fig 4B) [15, 16]. The data was consistent with the hit compound exhibiting specific and stoichiometric binding to HCV polymerase. Furthermore, the DLB data were reported that the left-hand portion of the hit compound was solvent exposed in the bound state and therefore was not in direct contact with a polymerase pocket or surface. Thus, subsequent analogues were designed, without the left-side segment. These follow-up analogues maintained the desired activity despite removal of the left-side, thus maintaining the essential benzimidazole core and right-side was consistent with DLB data suggesting that the latter both bound directly to polymerase [15, 16].

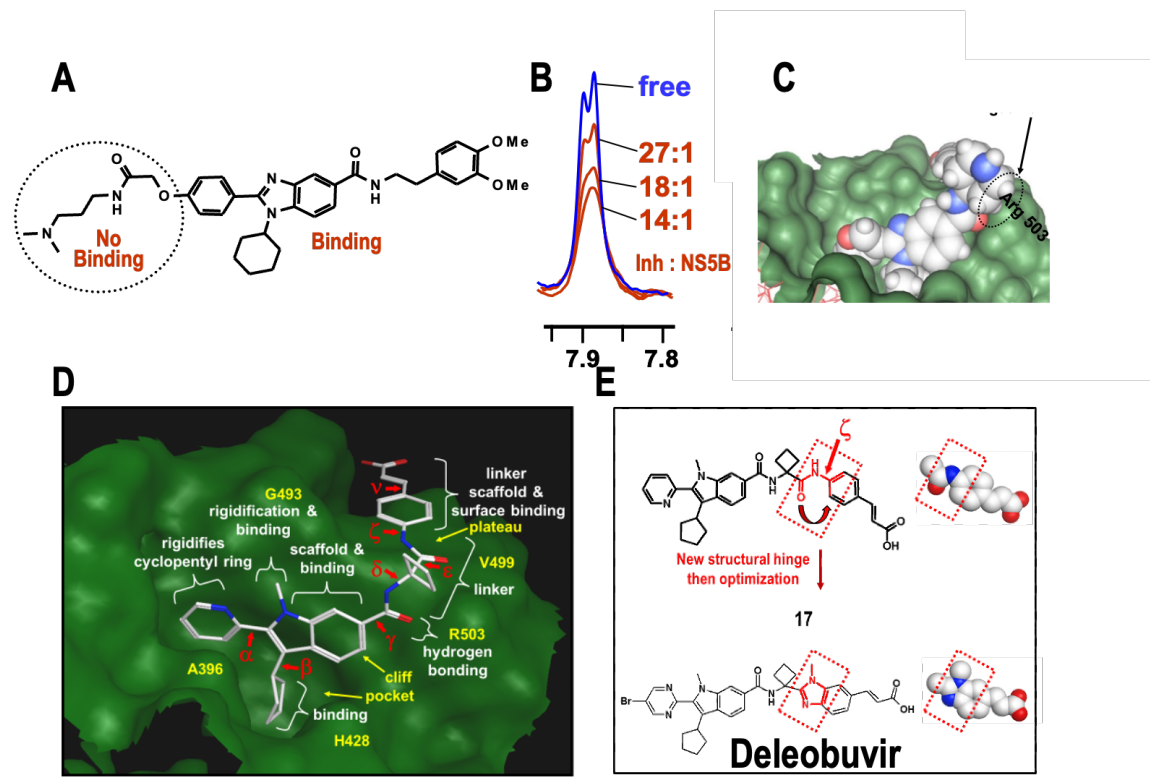

**Figure 4**. Targeting HCV polymerase. (A) Hit compounds from an HTS screen and validated by an NMR assay. (B) $^1$H NMR DLB data of free hit (blue) and after adding amounts of HCV polymerase

(red). (C and D) Docked NMR-derived structures of advanced inhibitors to pocket I of an X-ray structure of HCV polymerase. (E) Example of the design of Deleobuvir.

Transferred NOESY experiments were then applied to solve the bioactive bound conformations of more advanced analogues as shown in Fig 4C. Once determined, the bioactive structures were then docked to the pocket 1 site of the apo X-ray structure to produce the complexes shown in Fig. 4C and 4D. These complexes clearly showed that the ligand cyclohexyl group indeed bound into a deep pocket whereas the benzimidazole flat core tightly snugly clicked, like a lego piece, against a steep hydrophobic cliff partly formed by a proline side-chain, and the diamide linker helped the whole righthand segment to reach above the cliff edge and lie securely atop a complementary plateau of the polymerase. The roles of each segment of this series of inhibitors became clear (Fig. 4D), via multidisciplinary approaches, and design ideas became abundant. One example was depicted in Fig. 4E where NMR studies showed that analogues experienced dynamic flexibility of the structural hinge region in the free state (ROESY data), whereas only a single structure was observed for the bioactive bound conformation (transferred NOESY data). This observation and idea to match the free and bound states was termed "dynamics matching". This was accomplished by inserting a structural isostere (bottom compound in Fig. 4E), and the resultant series led to the clinical candidate Deleobuvir [14-16]. This work clearly showed how NMR can be central for opening new avenues within drug discovery projects, and can be valuable when integrated within multidisciplinary efforts.

## 3. Encountering issues from the phenomenon of compound self-association which led to new NMR detection and screening strategies

As mentioned above, when the HTS screen involving the HCV polymerase assay was executed [12,13], there were many issues as a result of compound self-aggregation into nano-entities and colloids. This natural phenomenon of compounds was partially responsible for the observation of high hit-rates and false-positives [17-19], and resulted in serious issues throughout the HCV polymerase program [21,22]. Although the phenomenon was initially observed and characterized during this program around the year 1997, it was and remains a major issue that requires constant attention in all small-molecule drug discovery programs.

A brief explanation of this strange phenomena is merited [20-23]. As indicated in Fig. 5, it was generally assumed that once a compound was placed in aqueous solution, it typically assumed a two-state existence. The compound could simply dissolve and adopt soluble fast-tumbling lone molecules in solution as shown in Fig. 5A, which could be detected by the observation of sharp NMR resonances. On the other hand, the compound could simply have limited solubility and result in some undissolved solid precipitate as shown over at Fig. 5D, which would result in the absence of solution NMR resonances. We and others clearly demonstrated [17-23], however, that poorly understood intermediate states also existed such as colloidal aggregates and nano-entities of various sizes (Figs. 5B and 5C). These slower tumbling self-assemblies could adopt a range of sizes. One of the reasons these entities were been poorly characterized was the lack of appropriate experimental detection methods. Thus, new detection methodologies had to be developed [20-21].

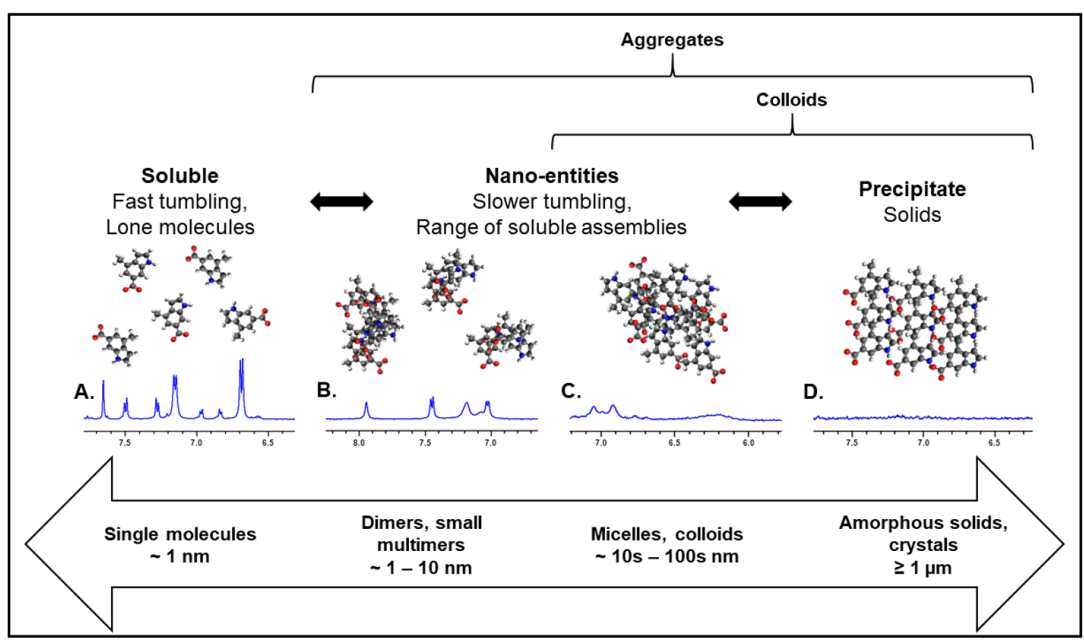

**Figure 5**. Phenomenon of compound aggregation. Shown is the three-phase model of the solution behavior of compounds [20]. Descriptions, sizes, scales and NMR spectra are given for each phase.

With regards to the HCV polymerase program, the phenomenon of nano-entities not only impacted the HTS triage and validation steps. It also had impacts throughout the whole program, which included the hit-to-lead and lead optimization workflows. NMR-based assays had to be developed to support this and other drug discovery programs. Thus, assays were applied throughout these workflows [21], and many findings were reported. One is that the NMR assay could qualitatively predict compound toxicity in off-target pharmacology assays [22], therefore, the assay was employed to screen advanced compounds and promote or deprioritize them for more advanced pre-clinical trial characterizations. Another report showed a correlation between the existence of nano-entities and immune responses [23].

**4. Design of antiviral drugs that target HIV**

4.1. Targeting HIV integrase and also encountering and resolving issues from the phenomenon of compound atropisomer chirality

The discovery of HIV antivirals requires one to think about the emergence of viral resistance, and thus the identification of inhibitors against multiple HIV protein targets was considered prudent. One attractive HIV protein was the integrase as it was essential for replication of the virus. A biochemical assay was developed with full-length integrase and adapted for HTS on over 1 million compounds [24-26]. Counter screens were implemented, and this helped to reduce the number of hits. Again, NMR assays were successfully employed to triage and validate HTS hits involving full-length integrase. NMR experiments were also practical for identifying exactly which subunits of integrase hits were binding to.

For this program, NMR played another critical and unexpected role with the discovery that the more advanced compounds adopted hindered internal rotation about a specific bond which gave rise to atropisomer axial chirality [27-29]. This phenomenon was first detected by 1D $^1$H NMR experiments for the only series that was being pursued for hit-to-lead purposes. In fact, NMR was the only practical method for detecting and characterizing this phenomenon, and so all new analogues synthesized for this program had to be characterized via NMR to determine their exact chirality (Fig. 6A). Fig. 6 displays example compounds where the barrier to rotation of the upper aromatic rings was hindered due to steric clashes with the bottom aromatic scaffold. Thus, this hindered rotation created axial chirality such that the upper rings can only adopt a left or right position with respect to the bottom scaffold (bottom panel of Fig 6A).

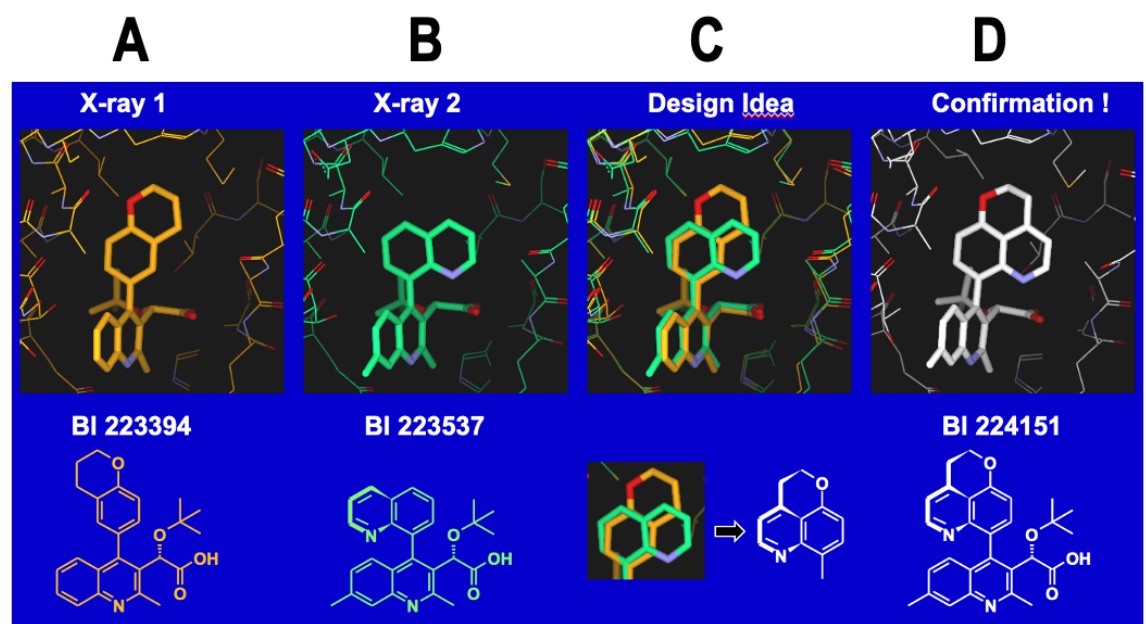

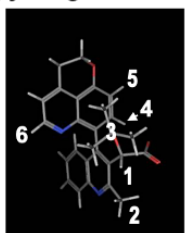
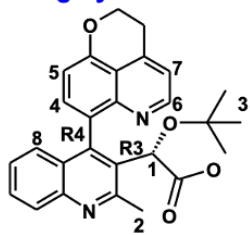
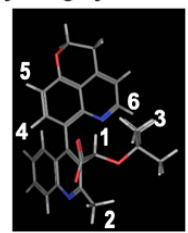

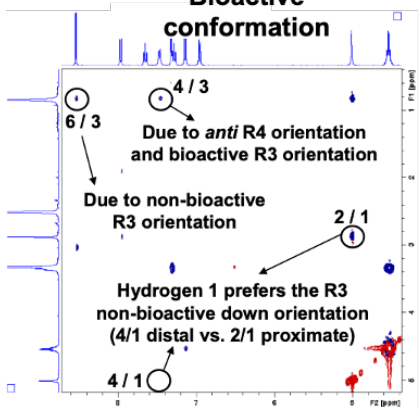
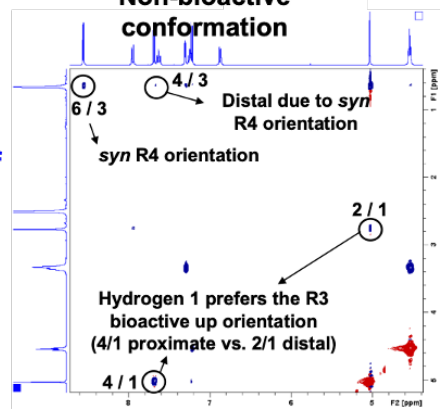

**Figure 6**. Targeting HIV integrase. (A) and (B) Shown are X-ray structures of two distinct atropisomers (confirmed by NMR), (C) along with the overlap and design idea, (D) followed by the synthesis and confirmation by X-ray and NMR. (E) Subsequent synthesis and NMR characterization of the clinical candidate BI 224436, and (F) its atopisomeric analogue. Detailed NMR ROESY characterization data is shown for both atropiosmers. For example, the larger the ROESY crosspeak, the closer are the hydrogens in the free-state.

Although this presented some workflow limitations, fruitful discoveries also resulted [24-26]. The rigidified atropisomers were more potent and selective, and allowed for design ideas. Once NMR was used to identify the correct chirality of potent atropisomeric compounds in the free state, selected compounds were submitted for X-ray crystallography to determine the 3D structures of the complexes. Here, we overlayed the 3D structure of the compounds shown in Fig 6A with that of the compounds shown in Fig. 6B, which then led to the idea of creating a top substituent that resembled the overlay shown in Fig. 6C. This led to the design of BI 224151 and eventually to the clinical candidate BI 224436, which required free-state characterization by NMR (Fig. 6E and 6F). This is yet another example of interdisciplinary synergy between NMR and crystallography and medicinal chemistry.

As this series of compounds evolved closer to the pre-development stage, questions arose about how one can ensure the delivery of a stable and consistent drug entity when epimerization changes can result by the simple rotation of a hindered bond. It turned out that the phenomenon of atropisomerism was also being experienced by most if not all major pharmaceutical companies. At that time, the common solution to this question was to simply abandon the compound series and seek alternate new chemical matter [27-29]. Fortunately, alternate options were sought for HIV series presented in Fig. 6. In fact, the whole phenomenon of atropisomerism and drug discovery was evaluated in depth, and a comprehensive review was published on the matter along with the FDA regulatory agency [27]. This successful review proposed options for moving compounds toward the clinic and has impacted many discovery campaigns. This and subsequent reports are highly cited in the literature [27-29].

4.2. Targeting HIV matrix

HIV matrix served as another reasonable target. Again, a biochemical assay was developed and subjected to a large library of compounds via an HTS screen. The use of subsequent counter screens was effective for significantly reducing the number of hits to those which were apparently more specific to the matrix biochemical assay.

Several NMR assays were subsequently deployed to help prioritize hits that attached directly to HIV matrix, and deprioritize those compounds that had unfavorable properties such as self-association into nano-entities. 1D $^1$H NMR data were very effective for this.

Practical NMR strategies were then used to address "deeper" questions that were relevant for this program. $^{15}$N HSQC NMR data served to monitor several features of interest:

specificity, direct binding to matrix, identification of the binding site on matrix, ligand binding affinities ($K_D$), and help to determine the mechanism of inhibition. For these experiments, $^{15}$N isotope-labelled matrix was expressed in *E.coli* with $^{15}$N enriched, minimal media, then purified and concentrated for NMR titration studies [29].

Fig. 7A shows a blue-colored $^1$H-$^{15}$N-HSQC dataset of unligated apo matrix where each peak arose from an amide H-N of each individual amino acid, thus the HSQC spectrum of peaks provides probes of each amino acid position. Overlayed atop of this blue HSQC spectrum of apo matrix was the red spectrum of matrix after adding a hit compound at a 1:10 ratio (Fig. 7A). Given that the red peaks overlay and hide the blue peaks, it is easy to conclude that the compound does not bind to and alter matrix. On the other hand, Fig. 7B showed an example of a hit compound that clearly bound to matrix. Upon titration of the compound to matrix, incremental changes were observed for specific HSQC peaks. Overall, these data were practical for confirming that this compound bound to matrix, that binding altered specific peaks and not others, the compound bound to the PIP2 site on matrix, and the titration data were used to determine the $K_D$ affinity of the ligand. Thus, chemical shift perturbations were very important to determine if ligands bind to the target and to determine the binding pocket.

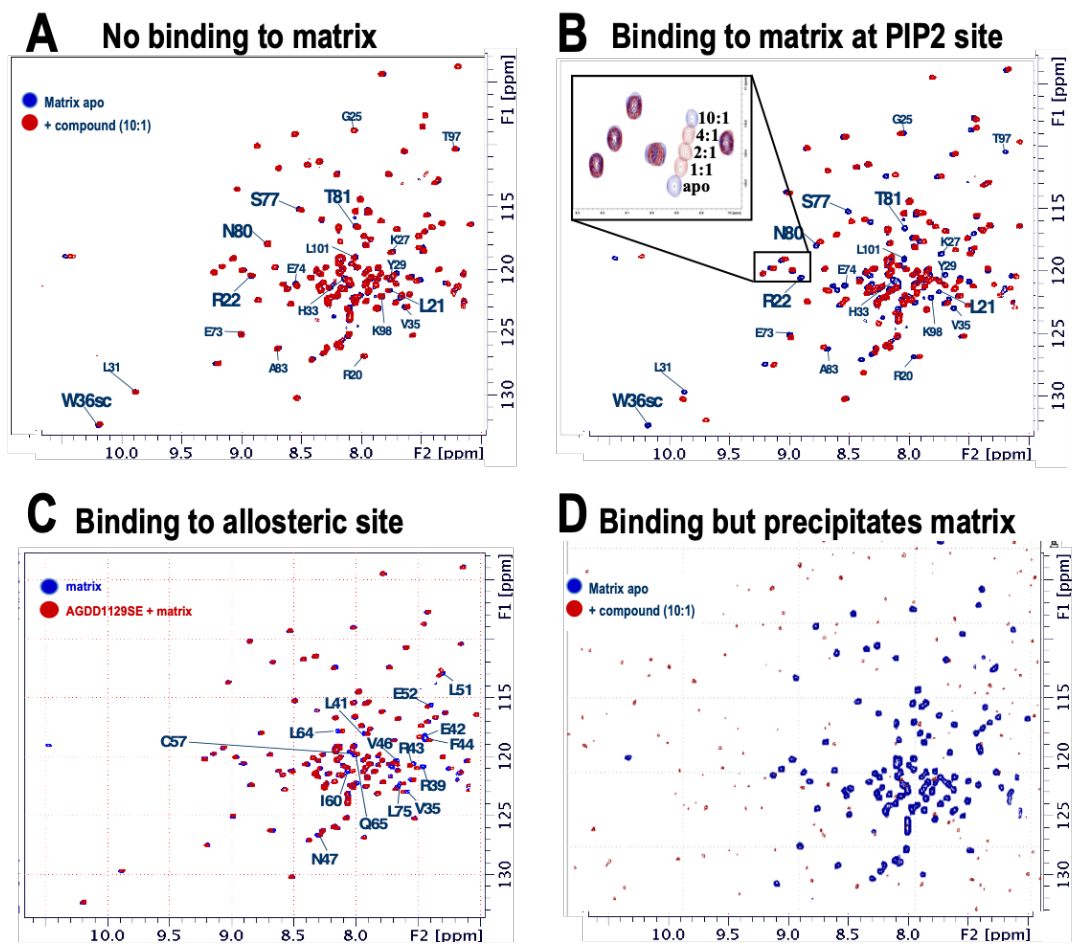

**Figure 7**. Targeting HIV matrix. (A-D) Shown are $^1$H-$^{15}$N HSQC NMR spectra of apo matrix protein (blue peaks) and matrix after adding four equivalents of the screening hit compounds (red).
In Figure 7B the magnification of the disturbed region does not correspond to that of the spectrum.

The results from the titration of another hit compound were shown in Fig. 7C. Again, specific peaks of matrix changed upon adding this hit compound, which confirmed binding to matrix at a specific site. However, as compared to the changes observed in Fig. 7B, the HSQC peaks of different amino acids actually changed upon binding to the matrix protein. Thus, it was confidently concluded that this hit compound bound to an allosteric site and not to the PIP2 site on matrix.

Finally, Fig. 7D provided a hint as to the mechanism of inhibition of another hit compound. When it was titrated to the matrix sample, a precipitant was observed, and no peaks were visible in the HSQC spectrum. It was likely that the compound demonstrated activity in assays as a result of precipitation of the complex, which may or not be a desired mechanism of action.

Ironically, the hit shown in 7B had exactly the same chemical structure of a compound from an active chemical series used to target pocket II of HCV polymerase. NMR and X-ray analyses of both clearly showed that one enantiomer bound to matrix whereas the opposite chiral enantiomer bound to HCV polymerase [45]. Of course, this was amazing, and it demonstrated the fact that atropisomer pairs are truly distinct compounds and should be treated that way.

4.3. Targeting HIV protease

Another attractive protein target was HIV protease, which belongs to the aspartyl protease enzyme family. The strategy employed to secure lead inhibitors was to mimic the P-P' peptidic sequence of the cleavage substrate (Fig. 8A). NMR data were engaged to monitor the free-state conformation and dynamics of analogues to find distinctions from the bioactive conformation. Close collaborations with advancing medicinal chemists and biochemists allowed for exploitation of these differences, and ideas served to help systematically design improved peptidomimetic analogues. This eventually resulted in the clinical candidate Palinavir displayed in Fig. 8B [30]. Note that a comparison of the original peptide sequence with the structure of Palinavir showed highly similar backbone chemical structures. Unfortunately, this compound was halted before the clinical stage for various reasons which included concerns about the scaleup of lengthy and complex syntheses. Other pharmaceutical companies nonetheless proceeded to the clinic and marketed with their own versions of HIV protease peptidomimetics. Interestingly, non-peptidic alternate versions of HIV protease inhibitors (Aptivus) were discovered as shown in Fig. 8C. This drug was impressively active against many resistant mutants of HIV protease.

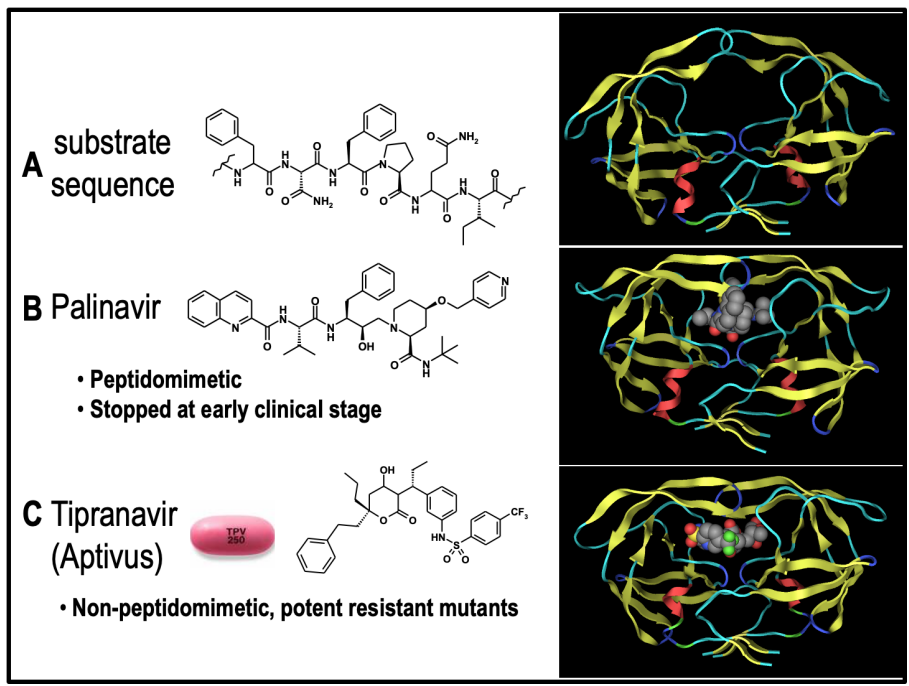

**Figure 8**. Targeting HIV protease. (A) Shown is the substrate amino acid sequence along with a view of the X-ray structure of apo HIV protease. Palinavir (B) and tipranavir (C) are shown with their respective X-ray structure with HIV protease.

4.4. Targeting HIV polymerase

HIV polymerase was another critical target for antiviral drug discovery. The strategies applied for drug discovery included both HTS and structure-based design [31-33]. One of the first HIV drugs to be discovered was Viramune (Fig. 9A). However, Viramune treatments in HIV patients soon resulted in the emergence of resistant mutants, especially Y181 and K103. See the panel on the right-top side of Fig. 9. Thus, new drug alternatives were urgently needed to combat the wild-type and resistant mutants. Structure-based design based on X-ray structures of complexes led to BILR0355 (Fig. 9B) and was found to inhibit wild type HIV and mutants (top-right panel of Fig. 9).

Efforts nonetheless continued to identify other alternative drugs. Another HTS campaign was launched followed by the established triaging and validation assays described above. Interestingly, an unusually potent hit was identified (Fig. 9C) and followed up with analogue optimizations. This included monitoring the free-state conformational attributes of compounds by NMR and computer-aided drug design (CADD), and making careful comparisons with the bound-state of compounds (determined by X-ray and CADD) (data not shown). Strategies such as "torsion angle dynamics matching" and scaffold hopping [33] were employed as rational design efforts. This work clearly showed how NMR can be valuable when integrated within multidisciplinary efforts.

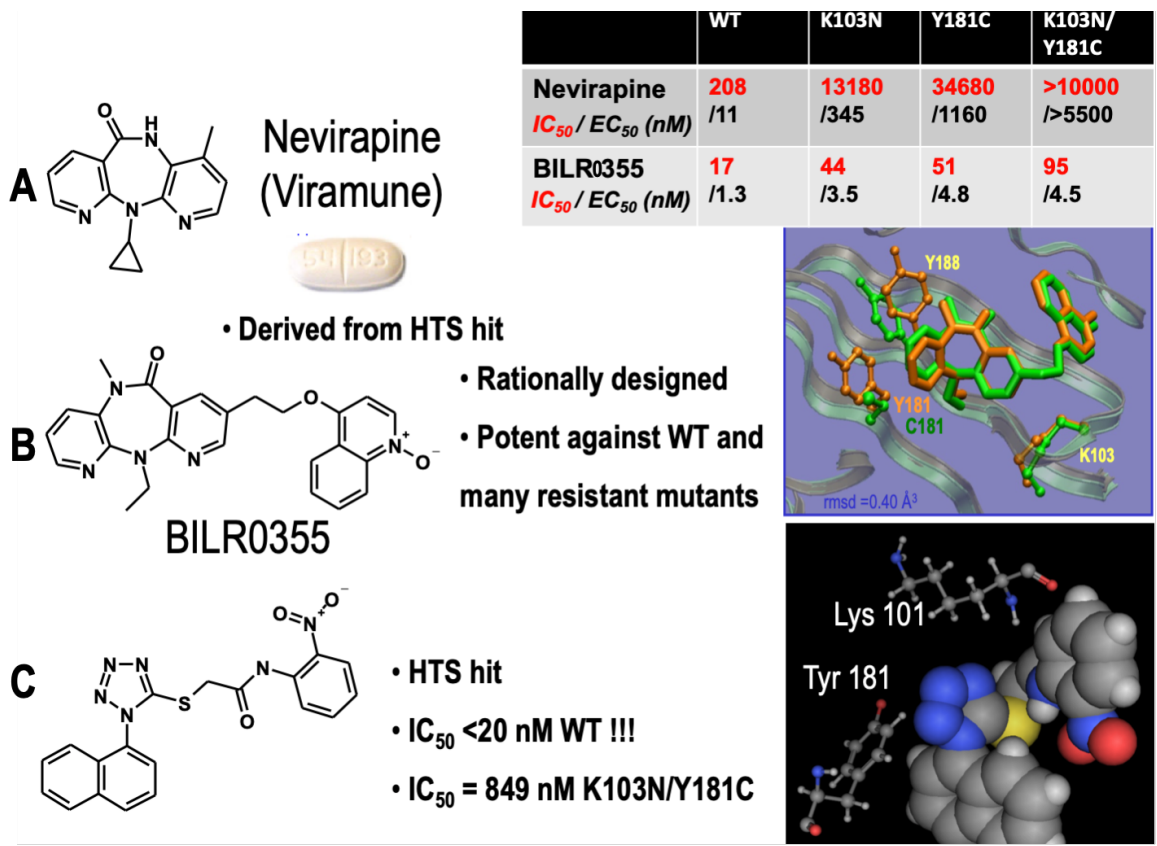

**Figure 9**. Targeting HIV polymerase. (A) Shown is Nevirapine along its activity profiles. (B) Shown is BILR0355 along with an X-ray structure of its complex with HIV polymerase. (C) Shown is an HTS hit along with zoom of an X-ray structure of its complex with HIV polymerase.

4.5. Targeting HIV maturation

Bevirimat (Fig. 10) was found to be an effective antiviral HIV drug that blocked maturation by affecting the cleavage of the capsid-spacer peptide 1 (CA-SP1) junction. Also, the related derivative EP-39 (Fig. 10), a more hydrophilic derivative, exhibited an interesting profile of activity and was expected to have similar antiviral attributes. As part of a perceptive multidisciplinary study [34-39], NMR and docking helped by providing deeper insights into the distinct mechanisms of actions of Bevirimat and EP-39. NMR data from the 1H-NOESY experiment (Fig. 10E,F) confirmed that the interaction of EP39 with a mutated peptide domain on SP1 (CA-SP1(A1V)-NC) was not detectable compared to the wild-type peptide.

This suggests that mutation of the first SP1 residue induces a loss of interaction between Pr55Gag and EP-39, as previously reported. These data make it possible to propose a mechanism of resistance of the virus to maturation inhibitors.

Moreover, EP-39 and Bevirimat (BVM) could interact differently with the Pr55Gag lattice. This was corroborated by the modeling results displayed in Figs. 10A,B,C,D and which proposed distinct binding modes for Bevirimat and EP-39. These combined data were also consistent with the finding that distinct mutants could be raised when exposed to EP-39 having four mutations within the CA domain (CA-A194T, CAT200N, CA-V230I, and CA-V230A) and one in the first residue of SP1 (SP1-A1V), versus the mutants raised when exposed to Bevirimat (CA-A194T, CA-L231F, CA-L231M, SP1-A1V, SP1-S5N and SP1-V7A). Together, these studies demonstrate the utility of NMR when used judiciously with multidisciplinary strategies [34-39].

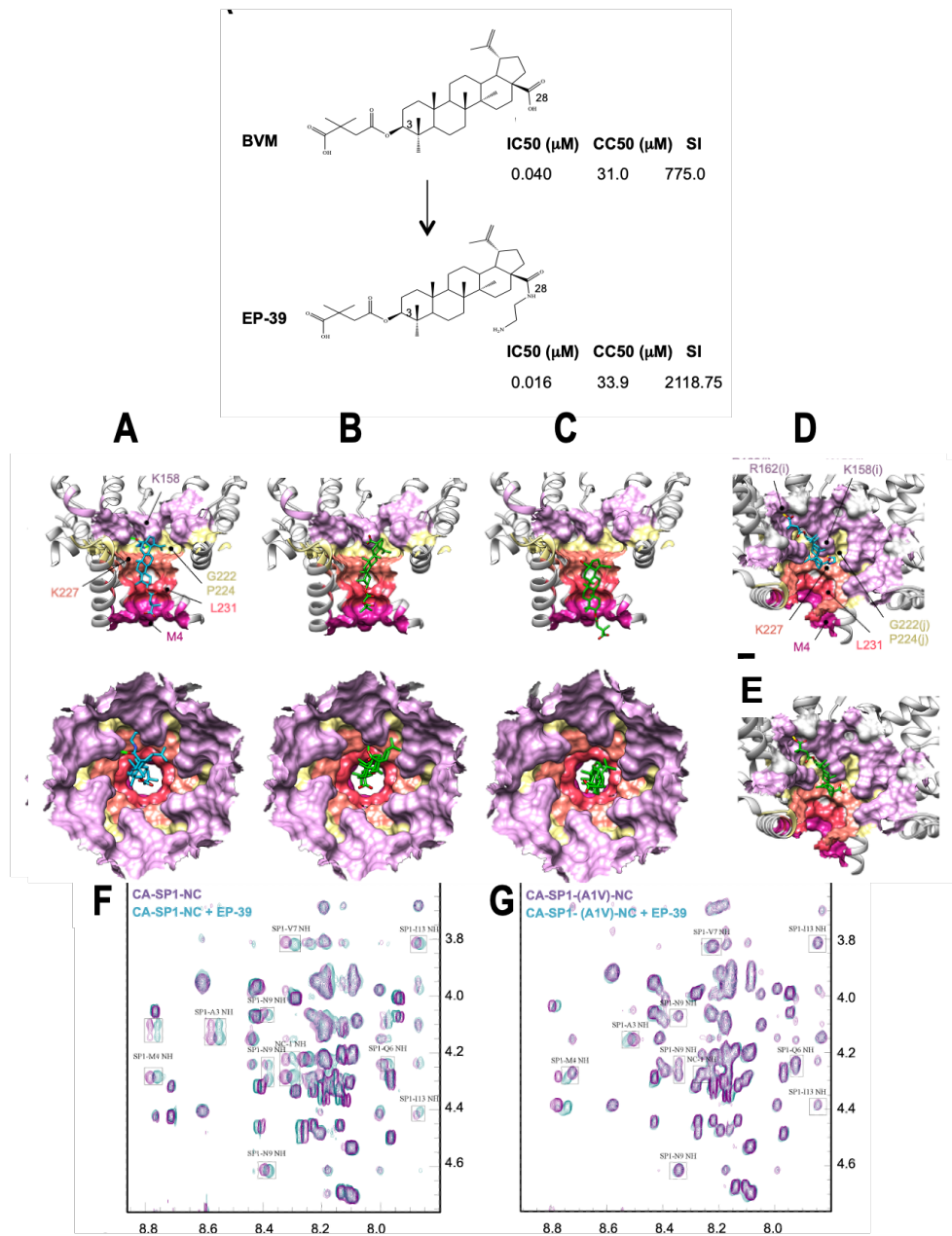

**Figure 10**. Targeting maturation. Shown on top are the chemical structures of Bevirimat (BVM) and EP-39. (A-E) In silico docking of EP-39 and Bevirimat on the hexameric crystal structure of the CACTD-SP1 Gag fragment. Views of the binding of EP-39 (A) and BVM (B) within the lower part of the barrel formed by the 6-helix bundle stem, where both are in contact with the six chains of the hexamer. An additional location, within the barrel, is observed for Bevirimat which can form new contacts (C). In (D - EP-39) and (E – Bevirimat) interact with the upper part of the hexamer and spread out along the interface. 2D $^1$H-NOESY NMR experiments of the interaction between EP-39 and WT CA-SP1-NC (F) and mutated CA-SP1(A1V)-NC (G).

## 5. Identifying antiviral drugs that target SARS-CoV-2 - Jumping from fragments to drugs

Drug repurposing can also be a value strategy for discovering interesting antiviral compounds. The concept is based on the idea that drugs already approved by regulatory agencies, and marketed for another indication, can potentially be reused for another purpose. The strategy has been used many times and takes advantage of the fact that small molecule drugs often bind to more than one target. A big advantage of drug repurposing is that these approved drugs have already passed safety and toxicity tests in clinical trials, and thus can advance more rapidly and cheaper in follow-up antiviral clinical phases.

With the onset of the COVID19 epidemic, it was highly desirable to explore drug repurposing strategies as a rapid means for discovering and using antivirals. However, serious hurdles were expected due to the demanding safety regulations required for screening SARS-CoV-2 (i.e. BSL3 laboratory). So, minimal testing experiments was desirable.

Thus, a new approach involving NMR fragment screening was implemented to pre-select compounds for antiviral testing [40]. The approach involved first screening a small library of fragment-like compounds to identify which chemical scaffolds bind to ACE-2 - the human receptor of the SARS-CoV-2 virus. Targeting human ACE-2 was considered prudent given that ACE-2 inhibitors would be expected to inhibit both wild-type and variant viruses.

Fig. 11A showed that an $^{19}$F-NMR fragment screen resulted in the discovery that fragment 14 binds to ACE-2. DLB and $^{19}$F-CPMG experiments confirm this binding. Note that the blue spectra of free fragment 14 were very different from those of fragment 14 in the presence of ACE-2 (red). Having identified a special binding scaffold to ACE-2, it was then appropriate to consider this as a "smart scaffold" (without the CF$_3$ group as shown in Fig 11B). This "smart scaffold" was then utilized to computationally search databases for structurally-related FDA-approved drugs. Fig. 11B displayed Vortioxetine which indeed was structurally very similar to fragment 14.

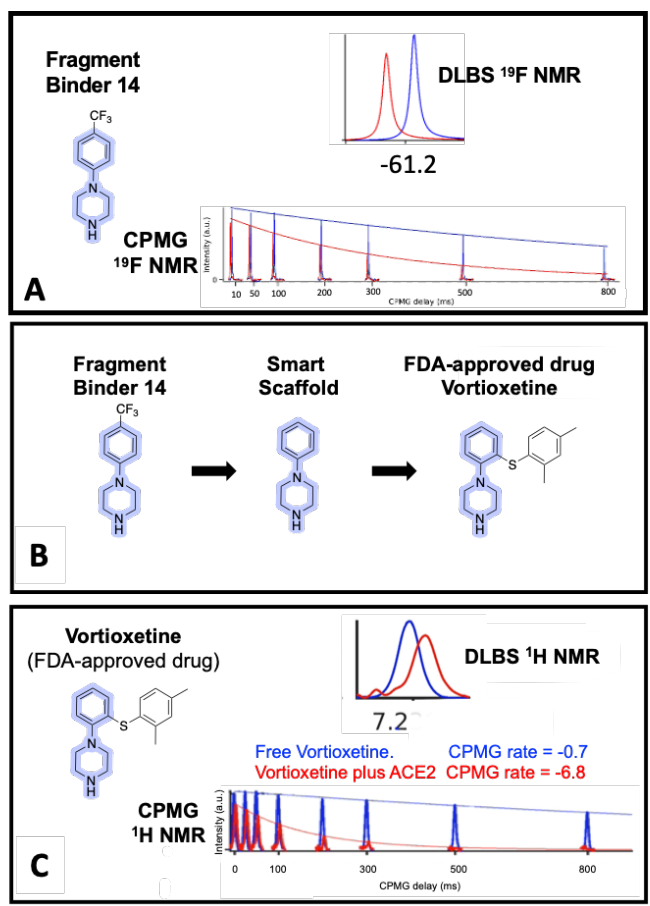

**Figure 11**. Targeting SAR-CoV-2. (A) Fragment 14 is displayed along with the NMR data used to detect it in a fragment screen. (B) Shown are Fragment 14, the derived "smart scaffold" and Vorioxetine. (C) Vortioxetine along with NMR data that demonstrated it binds to ACE-2.

Confirmation NMR studies were then run to determine whether or not Vortioxetine attached to ACE-2 as predicted. Fortunately, it was confirmed by DLB and CPMG experiments that Vortioxetine indeed attached directly to ACE-2. This was corroborated by ELISA tests which demonstrated that Vortioxetine inhibited the SARS-CoV-2 RBD binding domain from attaching to ACE-2 [40].

Moreover, Vortioxetine was tested under BSL3 laboratory conditions and shown to inhibit SARS-CoV-2 viral replication with an $EC_{50}=1.3$ μM. Although it was proposed that this compound has potential as an antiviral drug, more potent and patentable compounds are needed. For this, phenotype screens were run on a library collection of clinically-tested compounds and found potent compounds, along with combinations that resulted in synergic improvements in inhibitory activity against wild-type and variant SARS-CoV-2 (unpublished data). Combinations could also be more amenable for patentability, since many single compound activities have already been published in the rush to contribute to the anti-COVID19 effort. A future source of hit compounds could be derived from the well-proven method called "fragment phenotype lead discovery" which is discussed elsewhere [41, 42].

In conclusion, the main utility of the "smart scaffold" example given above, is its potential general applicability. Use of this methodology can be valuable to discover smart scaffolds that can then help to identify clinically-approved drugs (or other library collections) that will help to quickly jump from fragments to drugs.

## 6. Conclusions

This review demonstrated a variety of NMR applications which revealed compound and protein properties. Within multidisciplinary teams, these properties were shown to help to spawn and execute drug design ideas.

It is nonetheless important to provide a brief critical overview of some advantages and disadvantages of NMR in the drug discovery domain. Due to the length restrictions of this review, we focused only on a few selected viruses and examples. There are many more that include flu, RSV and herpes viruses such as cytomegalovirus. This review also only focuses on bio-NMR examples, whereas the most critical application is for daily primary structure verifications of synthetic compounds by medicinal chemists. NMR is also critical for primary structure elucidations, and the reader is referred to two publications in this regard [43, 44]. It is also noteworthy that NMR can help with drug pharmacology studies such as metabolite identification and with crystal form evaluations for formulation purposes using solid-state NMR. Note that there are many excellent examples of solid-state NMR applications in the literature.

NMR also has many disadvantages and thus alternate biophysical techniques should also be considered. For example, NMR is a relatively insensitive technique and thus significant amounts of sample are required. Often, this requires optimized protein expression and purification methods. When employing protein-detection experiments, there are practical size limits and for mid-sized and larger proteins isotope labeling is required. Unlike X-ray crystallography or cryo-EM, the determination of the 3D structures of proteins or protein-ligand complexes is arduous and time-consuming. Thus, it is often pragmatically recommended to prioritize other methods over NMR for this purpose when possible. Another disadvantage of NMR is that experiments are best designed to address a single question. Although this can also be considered as an advantage, multiple samples and significant time is involved in addressing multiple questions. Also, other biophysical techniques should be considered when possible to address questions that are onerous by NMR – e.g. $K_D$ determination of slow-exchange ligand-protein interactions.

Nonetheless, it is clear that NMR spectroscopy can play important roles to accelerate drug discovery within multidisciplinary teams. However, knowledge of the advantages and disadvantages of NMR applications is critical so that pragmatic applications can be performed in a timely fashion. It's no use to come up with data after teammates have moved on to other questions or projects. Drug discovery programs move forward fast in pharma. Another important aspect to keep in mind is that NMR applications have the potential of revealing uncharacterized or poorly characterized natural phenomena. The examples of atropisomer axial chirality and compound self-association are introduced here. Others certainly exist and are awaiting to be discovered.

## 7. Acknowledgements

The authors are grateful to the fantastic colleagues and teams at Boehringer Ingelheim for their input into the some of the work described herein. The authors are also appreciative to the funding agencies that supported the preparation of this review. FACS-Acuity, CQDM SynergiQc program, MEIE and Mitacs and Institut Pasteur. Special thanks for Université de Paris Cité for providing funding and support for the "Professeur Invité" program.